\documentclass[11pt]{article}
\usepackage{axodraw}
\usepackage{epsfig}
\usepackage{amsfonts}
\usepackage{amsmath}
\usepackage{bbm,bm}
\usepackage{cite}
\allowdisplaybreaks[1]

\input pix.sty

\newcommand{\bZ}{\widehat{Z}}

\renewcommand{\vec}[1]{{\bf #1}}

\newcommand{\alphas}{\alpha_{\rm s}}
\newcommand{\Nf}{N_{\rm f}}
\newcommand{\Nc}{N_{\rm c}}

\newcommand{\T}{\rmii{$T$}}
\newcommand{\CF}{C_\rmii{F}}
\newcommand{\gB}{g_\rmii{B}}
\newcommand{\mE}{m_\rmii{E}}

\newcommand{\mZ}{m_\rmii{$Z$}}

\newcommand{\gE}{g_\rmii{E}}

\newcommand{\gammaE}{\gamma_\rmii{E}}

\newcommand{\rmO}{{\mathcal{O}}}
\newcommand{\bmu}{\bar\mu}

\def\lsi{\raise0.3ex\hbox{$<$\kern-0.75em\raise-1.1ex\hbox{$\sim$}}}
\def\gsi{\raise0.3ex\hbox{$>$\kern-0.75em\raise-1.1ex\hbox{$\sim$}}}

\newcommand{\gsim}{\mathop{\gsi}}

\newcommand{\nF}{n_\rmii{F}}
\newcommand{\nB}{n_\rmii{B}}
\newcommand{\rmii}[1]{{\mbox{\tiny\rm{#1}}}}

\newcommand{\re}{\mathop{\mbox{Re}}}

\newcommand{\Tint}[1]{{\hbox{$\sum$}\!\!\!\!\!\!\!\int\,}_{\!\!\!\!\raise-0.9ex\hbox{$\scriptstyle{#1}$}}}
\newcommand{\Tinti}[1]{{{\Sigma}\!\!\!\!\raise0.3ex\hbox{$\int$}_\rmii{${#1}$}}}
\newcommand{\Tintip}[1]{{{\Sigma'}\!\!\!\!\!\raise0.3ex\hbox{$\int$}_\rmii{${#1}$}}}

\newcommand{\bi}{\begin{itemize}}
\newcommand{\ei}{\end{itemize}}
\newcommand{\hide}[1]{ }

\newcommand{\deltabar}{\raise-0.02em\hbox{$\bar{}$}\hspace*{-0.8mm}{\delta}}
\def\TAsc(#1,#2)(#3,#4,#5)%
{\SetWidth{2.0}\CArc(#1,#2)(#3,#4,#5)\SetWidth{1.0}}
\def\Lwidth{3}

\def\TAgl(#1,#2)(#3,#4,#5){\SetWidth{2.0}\PhotonArc(#1,#2)(#3,#4,#5){\Lwidth}%
{6.283 #3 mul 360 div #4 #5 sub #4 #5 sub mul sqrt mul Tdensity mul}%
\SetWidth{1.0}}
\def\TLgl(#1,#2)(#3,#4){\SetWidth{2.0}\Photon(#1,#2)(#3,#4){\Lwidth}
{#1 #3 sub #1 #3 sub mul #2 #4 sub #2 #4 sub mul add sqrt Tdensity mul}%
\SetWidth{1.0}}
\def\Aegl(#1,#2)(#3,#4,#5){\PhotonArc(#1,#2)(#3,#4,#5){\Lwidth}
{6.283 #3 mul 360 div #4 #5 sub #4 #5 sub mul sqrt mul Ldensity mul}}
\def\Legl(#1,#2)(#3,#4){\Photon(#1,#2)(#3,#4){\Lwidth}
{#1 #3 sub #1 #3 sub mul #2 #4 sub #2 #4 sub mul add sqrt Ldensity mul}}
\def\ToprSBB(#1,#2,#3,#4,#5){\picb{#1(0,15)(7.5,15)  #1(37.5,15)(45,15)%
 #2(22.5,15)(15,0,70) #2(22.5,15)(15,110,180) #3(22.5,15)(15,180,360)%
 #4(22.5,30)(5,-10,190) #5(22.5,30)(5,190,350)}}
\def\ToprSBT(#1,#2,#3,#4){\picb{#1(0,15)(7.5,15)  #1(37.5,15)(45,15)%
 #2(22.5,15)(15,0,90) #2(22.5,15)(15,90,180) #3(22.5,15)(15,180,360)%
 #4(22.5,35)(5,-90,270)}}
\def\ToprSTB(#1,#2,#3,#4){\picb{#1(0,0)(22.5,0) #1(22.5,0)(45,0)%
 #2(22.5,15)(15,-90,70) #2(22.5,15)(15,110,270)%
 #3(22.5,30)(5,-10,190) #4(22.5,30)(5,190,350)}}
\def\ToprSTT(#1,#2,#3){\picb{#1(0,0)(22.5,0) #1(22.5,0)(45,0)%
 #2(22.5,15)(15,-90,90) #2(22.5,15)(15,90,270)%
 #3(22.5,35)(5,-90,270)}}

\renewcommand{\picb}[1]{\;\parbox[c]{45pt}{\begin{picture}(45,30)(0,0)
\SetWidth{1.0}\SetScale{0.7} #1 \end{picture}}\;}
\renewcommand{\picc}[1]{\;\parbox[c]{60pt}{\begin{picture}(60,30)(0,0)
\SetWidth{1.0}\SetScale{1.3} #1 \end{picture}}\; }
\def\Lwidth{1.3}

\makeatletter \@addtoreset{equation}{section} \makeatother
\renewcommand{\theequation}{\arabic{section}.\arabic{equation}}
\makeatletter
\renewcommand\section{\@startsection {section}{1}{\z@}%
                                   {-5.5ex \@plus -1ex \@minus -.2ex}
                                   {2.3ex \@plus.2ex}%
                                   {\normalfont\large\bfseries}}
\renewcommand\subsection{\@startsection{subsection}{2}{\z@}%
                                     {-3.25ex\@plus -1ex \@minus -.2ex}%
                                     {1.5ex \@plus .2ex}%
                                     {\normalfont\normalsize\bfseries}}
\renewcommand\thesection {\@arabic\c@section}
\renewcommand\thesubsection   {\thesection.\@arabic\c@subsection}
\renewcommand{\@seccntformat}[1]{%
\csname the#1\endcsname.\hspace{1.0em}}
\makeatother

\begin{document}

\flushbottom

\begin{titlepage}

\begin{flushright}
December 2019
\end{flushright}
\begin{centering}

\vfill

{\Large{\bf
  A QCD Debye mass in a broad temperature range
}} 

\vspace{0.8cm}

M.~Laine$^{\rm a}$, 
P.~Schicho$^{\rm a}$,
Y.~Schr\"oder$^{\rm b}$

\vspace{0.8cm}

$^\rmi{a}$%
{\em
 AEC, Institute for Theoretical Physics, University of Bern, \\[1mm] 
 Sidlerstrasse 5, 3012 Bern, Switzerland\\
}

\vspace*{0.3cm}

$^\rmi{b}$%
{\em
 Grupo de Cosmolog\'ia y Part\'iculas Elementales,
 Universidad del B\'io-B\'io, \\[1mm]
 Casilla 447, Chill\'an, Chile
}

\vspace*{0.8cm}

\mbox{\bf Abstract}
 
\end{centering}

\vspace*{0.3cm}
 
\noindent
The Debye mass sets a scale for the screening of static charges and 
the scattering of fast charges within a gauge plasma. Inspired by its
potential cosmological applications, we determine a QCD Debye mass
at 2-loop order in a broad temperature range (1~GeV~...~10~TeV), 
demonstrating how quark mass thresholds get smoothly crossed. 
Along the way, integration-by-parts identities pertinent to
massive loops at finite temperature are illuminated.

\vfill

 
  
\vfill

\end{titlepage}

%
\section{Introduction} 
\la{se:intro}

If two test charges are put a distance $r$ apart
within a plasma, they influence
each other with a force which is weaker than 
the Coulomb force in vacuum, as a result of the 
screening caused by the light plasma particles. 
The potential then takes a Yukawa form, 
$- \alpha e^{-\mE^{} r}/r$, where 
the parameter $\mE^{ }$ may be called an electric or a Debye mass. 
In a relativistic plasma, it is of order $\mE^{ }\sim gT$, 
where $T$ is the temperature and $g$ is a gauge coupling. 

In the present paper we focus on strong interactions, such that 
$g$ is the coupling of the SU(3) gauge force. 
Standard applications of the 
QCD Debye mass can be found 
in the physics of heavy ion collision experiments. 
However, the temperatures 
reached there ($T \ll 1$~GeV) are so low that it is 
questionable whether perturbative tools are viable. 
Here we rather take $T \gsim 1$~GeV, and consider the possible 
role of the QCD Debye mass in cosmology. 

Given that strong interactions are in thermal equilibrium
in a broad temperature range, QCD does not normally
play a prominent role in cosmology. However, exceptions can be 
envisaged.\footnote{%
 None of the contexts listed here are ``urgent'',
 as they are related to yet-to-be-discovered BSM physics, nevertheless 
 we hope that put together they can motivate a well-defined QCD computation. 
 } 
For instance, it has become popular to 
consider scenarios in which dark matter is but the lightest 
among the particles of a larger dark sector. Then it is 
conceivable that the dark sector may also contain particles
charged under QCD (cf.,\ e.g,.\ ref.~\cite{garny} for a 
review of one such framework). 
At high temperatures, the pair annihilation 
of the QCD-charged particles would be modified by Debye
screening~\cite{threshold}.  
Charged particles also experience a thermal mass shift, 
known as the Salpeter correction in plasma physics, 
$\Delta M^{ }_\T \sim - \alpha\mE^{ }/2$, which can have 
a relatively speaking large effect if 
narrow degeneracies are present in the dark sector. 

Another possible
application concerns the decay of heavy particles, for instance
right-handed neutrinos in leptogenesis. 
In this case it is important
to know how fast the decay products 
(some of which could be hadronic, 
produced through the Higgs channel) 
equilibrate kinetically~\cite{bw}. 
This requires large-angle scattering, 
again sensitive to Debye screening~\cite{pa}.
Another relevant rate, namely 
that of decoherence of the decay products, originates from a 
difference of small-angle scatterings mediated by colour-magnetic
and colour-electric fields, whereby it is 
non-vanishing at $\rmO(\alpha T)$ thanks to $\mE^{ }\neq 0$
(cf.,\ e.g.,\ ref.~\cite{qhat}).

A third application of Debye masses is that they play a role
in dimensionally reduced descriptions of the electroweak phase 
transition~\cite{generic}. In particular, 
the QCD Debye mass could make a noticeable appearance
if some coloured scalar field is light enough to participate 
in the transition dynamics 
(cf.,\ e.g.,\ refs.~\cite{ewpt1,ewpt2}).

This paper is organized as follows. 
The definition of a Debye mass is subtle beyond leading order, 
so we start by specifying the concept adopted 
in \se\ref{se:formulation}. 
The main steps and methods of the computation are described 
in \se\ref{se:steps}, 
and results are presented
in \se\ref{se:results}.
We conclude in \se\ref{se:concl}, relegating 
the evaluation of massive 1-loop and 
2-loop sum-integrals into appendix~A.

%
\section{Formulation of the problem} 
\la{se:formulation}

As mentioned at the beginning of \se\ref{se:intro}, 
the leading-order definition of a Debye mass can be related 
to the Yukawa screening of a static potential or, equivalently, 
to the thermal mass that colour-electric fields obtain. 
In SU($\Nc^{ }$) gauge theory with $\Nf^{ }$ massless
fermions, the classic result reads~\cite{jk}
\be
 \mE^2 = g^2 T^2 \, \biggl(
  \frac{\Nc^{ }}{3} + \frac{\Nf^{ }}{6} \biggr)
 + \rmO(g^3 T^2)
 \;. 
\ee

When we go beyond leading order, the definition of a Debye mass
is no longer unique. One possibility is to define it as the 
inverse of a spatial correlation length related to some
gauge-invariant operator~\cite{ay}. This way, the Debye mass
becomes non-perturbative at next-to-leading order (NLO)~\cite{akr}. 
However, correlation lengths depend strongly on the 
quantum numbers of the operator chosen. There are also   
other non-perturbative possibilities, related e.g.\ to 
modelling the behaviour of the static potential at intermediate
distances~\cite{br}. 

A different strategy is to define the Debye mass as 
a ``matching coefficient'' of a low-energy description, 
specifically of a dimensionally reduced effective theory~\cite{dr1,dr2}.
There are a number of advantages with this strategy. One is that 
the definition is then ``universal'', with the same value 
appearing as an ingredient in the computation of 
many different correlation lengths~\cite{screening}, or even of dynamical
rates~\cite{qhat}. Another is that as a matching coefficient
$\mE^2$ is only sensitive to the hard scales 
that have been integrated out, and 
therefore perturbative by construction. In fact, the result
is known analytically up to 3-loop order in pure 
Yang-Mills theory~\cite{jm,gms}, and shows remarkable 
convergence down to low temperatures. 
Fermionic effects are for this definition currently known up to 2-loop
order in the massless limit~\cite{bn}, and up to 1-loop
level in the massive case~\cite{jk}. The purpose
of the current study is to extend the 2-loop result for $\mE^2$ 
to include massive fermions.\footnote{%
 We note in passing that 
 another analogous matching coefficient
 is the gauge coupling of the dimensionally reduced theory, 
 denoted by $\gE^2$, 
 but its determination is 
 technically more challenging. 
 For pure Yang-Mills theory results are 
 available up to 3-loop level but only in somewhat 
 incomplete numerical form~\cite{ig,ig_gE,soft_gE}.  
 Massless fermions have been included up 
 to 2-loop level~\cite{gE2}, mass effects
 up to 1-loop level~\cite{pheneos}.
}

To be explicit, 
the action of the dimensionally reduced effective theory, 
often called ``Electrostatic QCD'' (EQCD), reads
\be
 S^{ }_\rmii{EQCD} \; \equiv \;  
 \int_X \biggl\{  
  \fr14 F^a_{ij}F^a_{ij}
 +
  \fr12 \mathcal{D}^{ab}_i\! A^b_0\, \mathcal{D}^{ac}_i\! A^c_0 
 + 
  \frac{\mE^2}{2} A_0^a A_0^a 
 + 
 \ldots 
 \biggr\} 
 \;, 
\ee
where we are employing Euclidean conventions; 
$\int_X \equiv \frac{1}{T} \int \! {\rm d}^d\vec{x}$;  
$
 d = 3 - 2\epsilon
$;
$
 F^a_{ij} \equiv \partial^{ }_i A^a_j - \partial^{ }_j A^a_i
 + \gE^{ } f^{abc} A^b_i A^c_j 
$; 
$
 \mathcal{D}^{ab}_i \equiv 
 \delta^{ab} \partial_i - \gE^{ } f^{abc}A^c_i$; 
and $A_0^a$ is an adjoint scalar field.

In order to determine $\mE^2$, it is convenient to use the 
background field gauge~\cite{lfa} as a probe. We compute the temporal
2-point function with a purely spatial external momentum, 
\be
 \Pi^{ }_{00}(\vec{p}) 
 \equiv \Pi^{ }_\rmii{E}(p^2)
 = \sum_{n=1}^{\infty} \gB^{2n} \Pi^{ }_\rmii{E$n$}(p^2) 
 \;, \quad
 p \; \equiv \; |\vec{p}|
 \;. 
\ee
Here $\gB^{2}$ is the bare coupling, which is subsequently 
expressed in terms of the renormalized coupling $g^2$. 
The computation within full QCD (or the Standard Model) is matched
onto a computation within the effective theory, the latter also
re-expanded as a perturbative series in $g^2$. However, employing
dimensional regularization 
and Taylor-expanding in external momentum, 
the latter computation gives a vanishing result, given that
no scales appear in the propagators. Therefore, the
matching coefficient is directly given by a Taylor-expanded 
full theory computation, after accounting for different
field normalizations (or wave function corrections) within
the full and effective theories, 
\be
 \mE^2 = \gB^2 \, \Pi^{ }_\rmii{E1}(0) 
 + \gB^4 \, 
 \bigl[ \Pi^{ }_\rmii{E2}(0) -
 \Pi^{ }_\rmii{E1}(0) \Pi'_\rmii{E1}(0) \bigr]
  + \rmO(\gB^6)
 \;. \la{master}
\ee

%
\section{Main steps of the computation} 
\la{se:steps}

%
\begin{figure}[t]
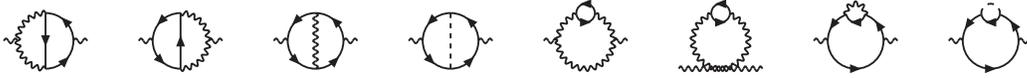


\hspace*{0.1cm}%
\begin{minipage}[c]{15.0cm}
\begin{eqnarray*}
&&{} \hspace*{-0.3cm}
 \ToptSM(\Legl,\Aqu,\Agl,\Agl,\Aqu,\Lqu)
 \ToptSM(\Legl,\Agl,\Aqu,\Aqu,\Agl,\Laqu)
 \ToptSM(\Legl,\Aqu,\Aqu,\Aqu,\Aqu,\Lgl)
 \ToptSM(\Legl,\Aqu,\Aqu,\Aqu,\Aqu,\Lsc)
 \ToprSBB(\Legl,\Agl,\Agl,\Aqu,\Aqu)
 \ToprSTB(\Legl,\Agl,\Aqu,\Aqu) 
 \ToprSBB(\Legl,\Aqu,\Aqu,\Agl,\Aaqu)
 \ToprSBB(\Legl,\Aqu,\Aqu,\Asc,\Aaqu)
\end{eqnarray*}
\end{minipage}

\vspace*{5mm}

\caption[a]{\small 
 Fermionic 2-loop contributions to the 
 gluon 2-point function. Wiggly lines denote gluons, solid lines
 quarks, and dashed lines Higgs bosons. 
}
\la{fig:graphs}
\end{figure}
%

The Feynman diagrams required for determining 
the 2-loop fermionic contributions
to $\mE^2$ are shown in \fig\ref{fig:graphs}
(we do not show gluonic diagrams as our results 
for them agree with ref.~\cite{bn}). Apart from vertices involving
the strong gauge coupling $g^2$, we have for illustration
also included effects from the 
top Yukawa coupling $h_t^2$, even if in practice these are small.\footnote{%
 For the latter set, only the scalar coupling to the physical Higgs mode 
 has been kept. The couplings to the Goldstone modes lead to 
 gauge-dependent contributions which can only be included if 
 the SU$^{ }_\rmii{L}$(2)$\times$U$^{ }_\rmii{Y}$(1) 
 gauge bosons are 
 incorporated as well, however 
 those effects are numerically very small. \label{yukawa}
}

The computation is carried out by employing the gauge propagator
\be
 \Bigl\langle A^a_\mu(P) A^b_\nu(Q)
 \Bigr\rangle 
 \; \equiv \;\, \deltabar(P+Q) \delta^{ab}
 \biggl[
 \frac{\delta^{ }_{\mu\nu}}{P^2} 
 + \frac{\xi P^{ }_\mu P^{ }_\nu}{P^4} 
 \biggr]
 \;,
\ee
where $\Tinti{P}\;\deltabar(P) = 1$ and 
$\Tinti{P} \equiv T\sum_{p^{ }_n} \mu^{3-d} \int \! 
\frac{{\rm d}^d\vec{p}}{(2\pi)^d}$ 
is a bosonic Matsubara sum-integral. 
We keep $\xi$ as a general gauge parameter (in the background
field gauge, it also appears
in the cubic and quartic gauge vertices~\cite{lfa}), 
verifying that it cancels exactly at the end. 

After carrying out the Wick contractions, the result can be expressed
in terms of the ``master'' sum-integrals 
\ba
 Z^{r}_{j;i} 
 & \equiv & 
 \Tint{\{P\}} \frac{p_n^r}{(P^2+m_i^2)^j}
 \;, \quad
 \bZ^{r}_{j;0} 
 \; \equiv \; 
 \Tint{P} \frac{p_n^r}{(P^2)^j} 
 \;, \la{Z1l} \\ 
 Z^{rs}_{jkl;i}
 & \equiv & 
 \Tint{\{P,Q\}}
 \frac{p_n^r q_n^s}{(P^2+m_i^2)^j (Q^2+m_i^2)^k [(P-Q)^2]^l}
 \;, \la{Z2l}
\ea
where $P \equiv (p^{ }_n,\vec{p})$ 
and $p^{ }_n$ is a Matsubara frequency. 
The sum-integral 
$\Tinti{\{P\}}$ signifies that $p^{ }_n$ is fermionic, i.e.\ 
$p^{ }_n \equiv \pi T (2n+1)$ with $n\in\mathbbm{Z}$.  
After renaming variables, 
the set of 2-loop masters can then be chosen to consist of 
$Z^{20}_{jkl;i}$, 
$Z^{11}_{jkl;i}$ and 
$Z^{ }_{jkl;i} \; \equiv \; Z^{00}_{jkl;i}$.
In rare cases, we add a mass $m^{ }_h$ on the bosonic
line and indicate this with the index $h$.
The mass index $i$ is omitted in intermediate results
if this can be done without a danger of confusion.

As far as 1-loop results are concerned, 
we need (cf.\ \eq\nr{master})
\ba
 \Pi^{ }_\rmii{E1}(0) & = & 
 (d-1)^2 \Nc^{ } \bZ^{ }_{1;0} 
 - 2 \sum_{i=1}^{\Nf}
 \bigl[
   (d-1) Z^{ }_{1;i} + 2m_i^2 Z^{ }_{2;i} 
 \bigr] 
 \;, \la{PiE1_1} \\ 
 \partial^{ }_{m_i^2}
 \Pi^{ }_\rmii{E1}(0) & = & 
  2 \sum_{i=1}^{\Nf}
 \bigl[
   (d-3) Z^{ }_{2;i} + 4m_i^2 Z^{ }_{3;i} 
 \bigr] 
 \;, \la{PiE1_2} \\ 
 \Pi'_\rmii{E1}(0) & = & 
 - \biggl[ \frac{d^2-5d+28}{6} - \xi (d-3) \biggr] \Nc^{ } \bZ^{ }_{2;0}  
 + \frac{1}{3} \sum_{i=1}^{\Nf}
 \bigl[
   (d-1) Z^{ }_{2;i} + 4 m_i^2 Z^{ }_{3;i} 
 \bigr] 
 \;. \la{PiE1_3} \hspace*{5mm}
\ea
Eq.~\nr{PiE1_2} is relevant because
$\Pi^{ }_\rmii{E1}$ is originally 
expressed as a function of bare
quark masses, which are expanded as 
$
 m^2_\rmii{B$i$}  =  
 m_i^2 \, 
 \bigl[ 
  1 + \frac{ 
    3 h_i^2 -12 g^2 \CF^{ }  
  }{2(4\pi)^2\epsilon}
   + \rmO(g^4)
 \bigr]
$, 
where $\CF^{ }\equiv (\Nc^2 - 1)/(2\Nc^{ })$.
In practice, Yukawa couplings $h_i^{ }$ other
than $h_t^{ }$ are omitted. 
Similarly, the bare gauge coupling is renormalized as 
$
 \gB^2  =  g^2 \, 
 \bigl[
   1 + \frac{g^2}{(4\pi)^2\epsilon}
   \bigl( 
     - \frac{11\Nc^{ }}{3} + 
       \frac{2}{3} \sum_{i=1}^{ \Nf^{ }}
   \bigr)
   + \rmO(g^4)
 \bigr]
$.

The 2-loop diagrams yield products of the 1-loop masters
of \eq\nr{Z1l} as well as genuine 2-loop masters defined
according to \eq\nr{Z2l}. All 
numerators can be eliminated from 1-loop masters by making use of 
$
 Z^{r+2}_{j+1;i} = 
 - m_i^2 Z^{r}_{j+1;i}
 + \bigl( 1 - \frac{d}{2j} \bigr) \, Z^{r}_{j;i}
$. 
This produces 
\ba
 \Pi^{ }_\rmii{E2}(0) & = & 
 -\Nc^2 (d-1)^2(d-3) (1-\xi) \bZ^{ }_{1;0} \bZ^{ }_{2;0}
 \nn
 & + & 
 \Nc^{ } \sum_{i=1}^{\Nf^{ }} 
 \Bigl\{ 
  2(d-1)(d-3)(1-\xi) Z^{ }_{1;i} \bZ^{ }_{2;0}
 + 4 m_i^2
 \bigl[
   Z^{ }_{112} 
  - \xi (d-3) Z^{ }_{2;i} \bZ^{ }_{2;0}
 \bigr]
 \nn
 & & \quad + \, 
 8 m_i^2 
 \bigl[
   Z^{11}_{221} 
  + 2 \bigl( Z^{11}_{212} - Z^{20}_{212} \bigr) 
  + 4 \bigl( Z^{11}_{113} - Z^{20}_{113} \bigr)
 \bigr]
 \Bigr\} 
 \nn
 & + & 
 \CF^{ } \sum_{i=1}^{\Nf^{ }}
 \Bigl\{ 
  2(d-1) \bigl[ \bZ^{ }_{1;0} - Z^{ }_{1;i} \bigr]
  \bigl[ (d-3) Z^{ }_{2;i} + 4 m_i^2 Z^{ }_{3;i} \bigr]
 \nn
 & & \quad + \, 
 8 m_i^2 
 \bigl[
   Z^{ }_{211} -2 Z^{11}_{221} - 4 Z^{20}_{311}  
 \bigr]
 \Bigr\} 
 \nn
 & - & \frac{ h_{t\rmii{B}}^2 }{\gB^2}
 \Bigl\{ 
 \bigl[ Z^{ }_{1;t} - \bZ^{ }_{1;h} \bigr] 
 \bigl[
   (d-3) Z^{ }_{2;t} + 4 m_t^2 Z^{ }_{3;t} 
 \bigr]
 \nn
 & & \quad + \, 
   \bigl( 4 m_t^2 - m_h^2 \bigr)
   \bigl[ Z^{ }_{211;th} - 2 Z^{11}_{221;th} - 4 Z^{20}_{311;th} \bigr]
 \Bigr\}
 \;. \la{PiE2_1}
\ea

The set of masters can now be reduced by making use of
integration-by-parts (IBP) identities~\cite{ibp1}, 
generalized to finite temperature~\cite{sn}. First, inspecting
\be
 0 = 
 \Tint{\{P,Q\}} \sum_{i=1}^d
 \frac{\partial}{\partial p^{ }_i}
 \frac{p^{ }_i \pm q^{ }_i}{(P^2+m^2)^j (Q^2+m^2)^k [(P-Q)^2]^l}
\ee
and taking linear combinations, leads to relations which permit us
to eliminate all quadratic powers of $p^{ }_n$, 
\ba
 Z^{20}_{(j+1)kl} & = & 
 \frac{1}{2j} \Bigl\{ 
   (2j+k-d) Z^{ }_{jkl} 
  - 2 m^2 \bigl[ j Z^{ }_{(j+1)kl} + k Z^{ }_{j(k+1)l} \bigr]
 \nn
 & & \quad + \, 
  k \bigl[
      Z^{ }_{(j-1)(k+1)l} - Z^{ }_{j(k+1)(l-1)} - 2 Z^{11}_{j(k+1)l}   
    \bigr]
 \Bigr\} 
 \;, \la{R1} \\ 
 Z^{20}_{jk(l+1)} & = & 
 \frac{1}{2l} \Bigl\{ 
    (l-k)Z^{ }_{jkl} + 2 k m^2 Z^{ }_{j(k+1)l} 
 \nn 
 & & \quad + \, 
 k \bigl[
     Z^{ }_{j(k+1)(l-1)} - Z^{ }_{(j-1)(k+1)l} + 2 Z^{11}_{j(k+1)l} 
   \bigr]
 \nn 
 & & \quad + \, 
 l \bigl[ 
     Z^{ }_{(j-1)k(l+1)} - Z^{ }_{j(k-1)(l+1)} + 2 Z^{11}_{jk(l+1)}
 \bigr]
 \Bigr\}
 \;. \la{R2}
\ea
Second, if we choose indices leading to two independent 
representations of some $Z^{20}_{jkl}$, we can establish relations
between $Z^{11}_{jkl}$. Considering $Z^{20}_{212}$
this way, we obtain from \eqs\nr{R1} and \nr{R2} the identity
\be
 2 \bigl( Z^{11}_{221} + 2 Z^{11}_{212} \bigr)
 = 
 Z^{ }_{2;i} \bigl( 2 \bZ^{ }_{2;0} - Z^{ }_{2;i}\bigr)
 - 2 m^2 \bigl( Z^{ }_{221} + 2 Z^{ }_{212} \bigr)
 - (d-2) Z^{ }_{112} 
 \;. \la{R3}
\ee
By using \eq\nr{R1} in order to eliminate $Z^{20}_{311}$, 
\eq\nr{R2} to eliminate $Z^{20}_{113}$ and $Z^{20}_{212}$, 
and inserting subsequently \eq\nr{R3}, we 
can remove all numerators from the sum-integrals of \eq\nr{PiE2_1}, leading to 
\ba
 \Pi^{ }_\rmii{E2}(0) & \supset & 
 \Nc^{ } \sum_{i=1}^{\Nf^{ }} 
 \Bigl\{ 
  2(d-1)(d-3)(1-\xi) Z^{ }_{1;i} \bZ^{ }_{2;0} 
 + 4 m_i^2
 \bigl[ 
  2 
  - \xi (d-3) 
 \bigr]
 Z^{ }_{2;i} \bZ^{ }_{2;0}
 \nn
 & & \quad + \, 
 4 m_i^2 
 \Bigl[
   (d-5) Z^{ }_{112} 
   - 2 m_i^2 Z^{ }_{221} 
   - \bigl( Z^{ }_{2;i} \bigr)^2
 \Bigr]
 \Bigr\} 
 \nn
 & + & 
 \CF^{ } \sum_{i=1}^{\Nf^{ }}
 \Bigl\{ 
  2(d-1) \bigl[ \bZ^{ }_{1;0} - Z^{ }_{1;i} \bigr]
  \bigl[ (d-3) Z^{ }_{2;i} + 4 m_i^2 Z^{ }_{3;i} \bigr]
 \nn
 & & \quad + \, 
 8 m_i^2 
 \Bigl[
     (d-5) Z^{ }_{211} 
   + 2 m_i^2 \bigl( Z^{ }_{221} + 2 Z^{ }_{311} \bigr) 
   + \bigl( Z^{ }_{2;1} \bigr)^2 
 \Bigr]
 \Bigr\}
 \;. \la{PiE2_2} 
\ea

Remarkably, IBP relations also exist between 
masters without any numerators. 
In this way, we can eliminate $Z^{ }_{221}$, 
$Z^{ }_{211}$ and $Z^{ }_{311}$ in favour of 
$Z^{ }_{111}$, $Z^{ }_{112}$ and $Z^{ }_{212}$. 
The latter set is convenient, as it turns out
that $Z^{ }_{111}$ appears with zero coefficient in $d$ dimensions, 
and $Z^{ }_{212}$ can be obtained from 
$Z^{ }_{112}$ through a mass derivative. Thereby
only one irreducible master, $Z^{ }_{112}$, remains
to be determined in detail (cf.\ appendix~\ref{ss:z112}).\footnote{%
 Taken on its own, the master $Z^{ }_{112}$ is IR divergent.
 However, the matching coefficient $\mE^2$ as a whole is IR safe
 by construction. For a proper cancellation of IR divergences, all masters
 need to be consistently evaluated with dimensional regularization,
 which regularizes both their IR and UV divergences.
}

The relations needed, originally found via our FORM~\cite{form}
implementation of Laporta-type reduction~\cite{ibp2},  read 
\ba
 Z^{ }_{211} & = & 
 - \frac{(d-3) Z^{ }_{111}}{4 m^2}
 + \frac{ ( Z^{ }_{2;i} )^2 }{2(d-2)}
 \;, \la{red1} \\
 Z^{ }_{221} + 2 Z^{ }_{311} & = & 
 \frac{(d-3)(d-5) Z^{ }_{111}}{8 m^4}
 + \frac{ Z^{ }_{2;i} 
  [ 8 m^2 Z^{ }_{3;i} -(d-3) Z^{ }_{2;i} ] }{4(d-2)m^2}
 \;, \la{red2} \\ 
 Z^{ }_{221} & = & 
 \frac{(d-2)(d-5) Z^{ }_{112}}{4 m^2}
 + (d-4) Z^{ }_{212}
 + \frac{Z^{ }_{2;i} ( 2 \bZ^{ }_{2;0} - Z^{ }_{2;i} ) }{4m^2}
 \;. \la{red3}
\ea
Defining
$
 \Delta^{ }_\rmii{$P$} \; \equiv \;  P^2 + m^2
$
and 
$
 \delta^{ }_\rmii{$P$} \; \equiv \; P^2 
$, 
\eqs\nr{red1} and \nr{red3} can be verified by setting 
$s=1$ and $s=2$, respectively, in the relation 
\ba
 0 & = & 
 \Tint{\{P,Q\}} \sum_{i=1}^d
 \frac{\partial}{\partial p^{ }_i}
 \biggl\{ 
  \frac{(d-2s)\, p^{ }_i}
       {\Delta^{ }_\rmii{$P$} \Delta^{ }_\rmii{$Q$} 
        \delta^{s}_\rmii{$P\!-\!Q$}}
 + 
  \frac{2 p^{ }_n (p^{ }_n \, q^{ }_i -  q^{ }_n\, p^{ }_i)}
       {\Delta^{ }_\rmii{$P$} \Delta^{2}_\rmii{$Q$} 
        \delta^{s}_\rmii{$P\!-\!Q$}}
 - 
  \frac{ p^{ }_i}
       {\Delta^{ }_\rmii{$P$} \Delta^{2}_\rmii{$Q$} 
        \delta^{s-1}_\rmii{$P\!-\!Q$}}
 +
  \frac{p^{ }_i - q^{ }_i}{\Delta^{2}_\rmii{$Q$}
        \delta^{s}_\rmii{$P\!-\!Q$}} 
 \biggr\} 
 \nn 
 & + & 
 \Tint{\{P,Q\}} \sum_{i=1}^d
 \frac{\partial}{\partial q^{ }_i}
 \biggl\{ 
  \frac{(d-2s)\, p^{ }_i}
       {\Delta^{ }_\rmii{$P$} \Delta^{ }_\rmii{$Q$} 
        \delta^{s}_\rmii{$P\!-\!Q$}}
 - 
  \frac{2 p^{ }_n (p^{ }_n \, q^{ }_i -  q^{ }_n\, p^{ }_i)}
       {\Delta^{2}_\rmii{$P$} \Delta^{ }_\rmii{$Q$} 
        \delta^{s}_\rmii{$P\!-\!Q$}}
 - 
  \frac{ (s-1)\, p^{ }_i}
       {\Delta^{2}_\rmii{$P$} \Delta^{ }_\rmii{$Q$} 
        \delta^{s-1}_\rmii{$P\!-\!Q$}}
 \biggr\} 
 \;, 
\ea
whereas \eq\nr{red2} can be established by taking a mass
derivative of \eq\nr{red1}.

Inserting \eqs\nr{red1}--\nr{red3}, the $\CF^{ }$ part gets factorized, 
and \eq\nr{PiE2_2} reduces to
\ba
 \Pi^{ }_\rmii{E2}(0)
 \!\!\! & \supset & \!\!\!  
 \Nc^{ }\sum_{i=1}^{\Nf^{ }}
 \Bigl\{ 
   2(d-1)(d-3) Z^{ }_{1;i} \bZ^{ }_{2;0} 
 + 2 m_i^2 Z^{ }_{2;i} 
   \bigl( 2 \bZ^{ }_{2;0} - Z^{ }_{2;i} \bigr) 
\nn & & \; 
 - \, 2\xi(d-3) \bZ^{ }_{2;0}[(d-1)Z^{ }_{1;i} + 2 m_i^2 Z^{ }_{2;i}] 
 - \, 2 (d-4) m_i^2 \bigl[ 
    (d-5) Z^{ }_{112}
    + 4 m_i^2 Z^{ }_{212} 
  \bigr]
 \Bigr\} 
 \nn 
\!\!\! & + & \!\!\! 
 \CF^{ }\sum_{i=1}^{\Nf^{ }}
  \bigl[ (d-3) Z^{ }_{2;i} + 4 m_i^2 Z^{ }_{3;i} \bigr]
 \biggl\{ 
   2(d-1)[\bZ^{ }_{1;0} - Z^{ }_{1;i}]
  +  \frac{8 m_i^2 Z^{ }_{2;i}}{d-2}
 \biggr\} 
 \;. \la{PiE2_3}
\ea

Adding to \eq\nr{PiE2_3}
the contributions from \eqs\nr{PiE1_1} and \nr{PiE1_3}
according to \eq\nr{master}, and re-installing the 
$\Nc^2$ and $h_t^2$ parts of \eq\nr{PiE2_1},  
we write the result in terms of $\msbar$-renormalized couplings as 
\ba
 \mE^2 & = &  g^2 \, \Bigl[ 
 \Nc^{ }\, \Phi^\rmii{(1)}_{ }
 + 
 \, \sum_{i=1}^{\Nf^{ }} \Phi^\rmii{(2)}_{i}
 \Bigr]
 \nn 
 & + & g^4 \, \Bigl[ 
      \Nc^2 \, \Phi^\rmii{(3)}_{ }
   +  \sum_{i=1}^{\Nf^{ }} \Bigl( \Nc^{ } \, \Phi^\rmii{(4)}_{i}
   +  \CF^{ } \, \Phi^\rmii{(5)}_{i} \Bigr)
   +  \sum_{i,j=1}^{\Nf^{ }} \Phi^\rmii{(2)}_{i} \Phi^\rmii{(6)}_{j}
  \Bigr]
 + g^2 h_t^2\, \Phi^\rmii{(7)}_{ }
 + \rmO(g^6) 
 \;. \hspace*{8mm} \la{Phi_def}
\ea
The various functions employed in \eq\nr{Phi_def} read
\ba
 \Phi^\rmii{(1)}_{ } & = & 
 (d-1)^2 \Nc^{ } \bZ^{ }_{1;0}
 \;, 
 \la{Phi1}
 \\
 \Phi^\rmii{(2)}_{i} & = & 
 -2\, 
 \bigl[
   (d-1) Z^{ }_{1;i} + 2m_i^2 Z^{ }_{2;i} 
 \bigr] 
 \;,
 \la{Phi2}
 \\
 \Phi^\rmii{(3)}_{ } & = & (d-1)^2 \bZ^{ }_{1;0} \, 
 \biggl[
 \frac{(d^{\,2} - 11d + 46)\,\bZ^{ }_{2;0}}{6} 
 - \frac{11}{(4\pi)^2 3\epsilon} \biggr]
 \;, 
 \la{Phi3}
 \\ 
 \Phi^\rmii{(4)}_{i} & = & 
 \bigl[(d-1) Z^{ }_{1;i} + 2 m_i^2 Z^{ }_{2;i} \bigr]
 \biggl[
   \frac{22}{(4\pi)^2 3\epsilon} 
  - \frac{(d^{\,2} - 5d + 28)\,\bZ^{ }_{2;0}}{3}
 \biggr]
 \nn 
 & + &  
 (d-1)^2 \bZ^{ }_{1;0} 
 \biggl[ 
    \frac{2}{(4\pi)^2 3\epsilon}      
   - 
    \frac{ (d-1) Z^{ }_{2;i} + 4 m_i^2 Z^{ }_{3;i} }{3}
 \biggr] 
 \nn 
 & + & 
   2(d-1)(d-3) Z^{ }_{1;i} \bZ^{ }_{2;0} 
 + 
   2 m_i^2 Z^{ }_{2;i} (2  \bZ^{ }_{2;0} - Z^{ }_{2;i})  
 \nn 
 & - & 
   2 m_i^2 (d-4)
   \bigl[ 
    (d-5) Z^{ }_{112}
   + 
    4 m_i^2 Z^{ }_{212}
   \bigr]
 \;, 
 \la{Phi4} 
 \\ 
 \Phi^\rmii{(5)}_{i} & = & \!\!
 - 
 \bigl[
        (d-3) Z^{ }_{2;i} + 4 m_i^2 Z^{ }_{3;i} 
 \bigr]
 \biggl[ 
 \biggl( 
    \frac{12}{(4\pi)^2 \epsilon} - \frac{8 Z^{ }_{2;i}}{d-2}
 \biggr) m_i^2
 + 2(d-1) \bigl( Z^{ }_{1;i} - \bZ^{ }_{1;0}\bigr)
 \biggr]
 \;,
 \la{Phi5} \hspace*{5mm}
 \\ 
 \Phi^\rmii{(6)}_{j} & = & 
 - \frac{1}{3} 
 \biggl[
   (d-1) Z^{ }_{2;j} + 4 m_j^2 Z^{ }_{3;j} 
  - \frac{2}{(4\pi)^2\epsilon}
 \biggr] 
 \,,
 \la{Phi6}
 \\ 
 \Phi^\rmii{(7)}_{ } 
  &
    \hspace*{-0.4cm}
    \stackrel{m^{ }_h \!\ll m^{ }_t}{\approx}
    \hspace*{-0.4cm} 
  & 
 - 
 \bigl[
        (d-3) Z^{ }_{2;t} + 4 m_t^2 Z^{ }_{3;t} 
 \bigr]
 \biggl[ 
 \biggl( 
    \frac{4 Z^{ }_{2;t}}{d-2} - \frac{3}{(4\pi)^2 \epsilon} 
 \biggr) m_t^2
 + Z^{ }_{1;t} - \bZ^{ }_{1;0}
 \biggr]
 \;.
 \la{Phi7}
\ea
These are gauge-independent, i.e.\ no $\xi$ appears, and also 
finite after the insertion of the masters from appendix~A, 
i.e.\ no $1/\epsilon^2$ or $1/\epsilon$ appears. 
For $ \Phi^\rmii{(7)}_{ } $ we have approximated 
the result by considering the limit $m^{ }_h \ll m^{ }_t$, as this leads 
to the same basis as for the pure QCD contributions. This  
overestimates the magnitude of $ \Phi^\rmii{(7)}_{ } $, 
but given that its effect is small even then
(cf.\ \fig\ref{fig:mmEoT2}), 
the approximation can be considered conservative. 

%
\section{Results} 
\la{se:results}

Our final results for the coefficients in 
\eq\nr{Phi_def} are obtained by inserting the master
sum-integrals from appendix~A into \eqs\nr{Phi1}--\nr{Phi7}  
and by then expanding the expressions up to $\rmO(\epsilon^0)$. 
Denoting
$
 \int_p \equiv \int\!\frac{{\rm d}^3\vec{p}}{(2\pi)^3}
$, 
$
 \omega^{ }_{p_i} \equiv \sqrt{p^2 + m_i^2}
$
and 
$
 \nF^{ }(\omega) \equiv 1/[\exp(\omega/T) + 1]
$,  
we find
\ba
 \Phi^\rmii{(1)}_{ } & \stackrel{d=3-2\epsilon}{=} &
 \frac{T^2}{3}
 \;, \la{Phi1_res} \\[2mm]
 \Phi^\rmii{(2)}_{i} & \stackrel{d=3-2\epsilon}{=} & 
 2 
 \int_p \frac{\nF^{ }(\omega^{ }_{p_i})}{\omega^{ }_{p_i}}
 \biggl( 2 + \frac{m_i^2}{p^2} \biggr)
 \;, \la{Phi2_res} \\[2mm] 
 \Phi^\rmii{(3)}_{ } & \stackrel{d=3-2\epsilon}{=} &
 \frac{22 T^2}{9(4\pi)^2}
 \, \biggl[ \ln\biggl(
   \frac{\bmu e^{\gammaE}}{4\pi T}
 \biggr) + \frac{5}{22} \biggr]
 \;, \la{Phi3_res} \\[2mm]
 \Phi^\rmii{(4)}_{i} & \stackrel{d=3-2\epsilon}{=} & 
 \frac{T^2}{9}
 \biggl[ 
  \int_p \frac{\nF^{ }(\omega^{ }_{p_i})}{p^2 \omega^{ }_{p_i}}
 - \frac{2}{(4\pi)^2} \ln\biggl(\frac{\bmu^2}{m_i^2}\biggr)
 \biggr]
 \la{Phi4_res} \\ 
 & & \; + \, 
 \frac{44}{3(4\pi)^2}
 \biggl[ 
    \ln\biggl( \frac{\bmu e^{\gammaE}}{4\pi T}
    \biggr)
    + \frac{1}{2} 
 \biggr]
 \int_p \frac{\nF^{ }(\omega^{ }_{p_i})}{\omega^{ }_{p_i}}
 \biggl( 2 + \frac{m_i^2}{p^2} \biggr)
 - \frac{8}{(4\pi)^2}
  \int_p \frac{\nF^{ }(\omega^{ }_{p_i})}{\omega^{ }_{p_i}}
 \nn 
 & & \; + \,
   \frac{m_i^2T^2}{18}  
   \biggl\{ 
  \int_p \frac{\nF^{ }(\omega^{ }_{p_i})}{p^2 \omega^{3}_{p_i}}
  + 
  \frac{1}{T}
  \int_p \frac{\nF^{ }(\omega^{ }_{p_i})
  [ 1 - \nF^{ }(\omega^{ }_{p_i}) ] }{p^2 \omega^{2}_{p_i}}
 \biggr\} 
 \; - \; 
 \frac{m_i^2}{2}\biggl\{ 
 \int_p \! \frac{\nF^{ }(\omega^{ }_{p_i})}{p^2 \omega^{ }_{p_i}}
 \biggr\}^2  
 \nn 
 & & \; + \, 
 \int_{p,q} 
 \frac{m_i^2}
 {8 \omega^{ }_{p_i} \omega^{ }_{q_i}} 
 \biggl( 
  \frac{1}{p^2}
 + 
  \frac{1}{q^2}
 \biggr)
 \biggl\{ 
   \frac{
    [ \nF^{ }(\omega^{ }_{p_i}) + \nF^{ }(\omega^{ }_{q_i}) ]^2
        }{
    ( \omega^{ }_{p_i} + \omega^{ }_{q_i} )^2
        }
   - 
   \frac{
    [ \nF^{ }(\omega^{ }_{p_i}) - \nF^{ }(\omega^{ }_{q_i}) ]^2
        }{
    ( \omega^{ }_{p_i} - \omega^{ }_{q_i} )^2
        }
 \biggr\}
 \;,   \nonumber  \hspace*{4mm} \\[2mm] 
 \Phi^\rmii{(5)}_{i} & \stackrel{d=3-2\epsilon}{=} & 
 - \frac{ m_i^2 }{2} \biggl\{ 
  \int_p \frac{\nF^{ }(\omega^{ }_{p_i})}{p^2 \omega^{3}_{p_i}}
  + 
  \frac{1}{T}
  \int_p \frac{\nF^{ }(\omega^{ }_{p_i})
  [ 1 - \nF^{ }(\omega^{ }_{p_i}) ] }{p^2 \omega^{2}_{p_i}}
 \biggr\} 
 \la{Phi5_res} \\ 
 &  & \; \times \, 
 \biggl\{
  \frac{T^2}{3} 
  + \frac{12 m_i^2}{(4\pi)^2}
  \biggl[ 
      \ln\biggl(\frac{\bmu^2}{m_i^2}\biggr) 
    + \frac{4}{3} 
  \biggr]
 + 
 4 \int_q \frac{\nF^{ }(\omega^{ }_{q_i})}{\omega^{ }_{q_i}}
 \biggl( 1 - \frac{m_i^2}{q^2} \biggr)
 \biggr\}
 \;, \nonumber \hspace*{4mm} \\[2mm] 
 \Phi^\rmii{(6)}_{j} & \stackrel{d=3-2\epsilon}{=} & 
 \frac{1}{3} 
 \biggl\{
  \int_q \frac{\nF^{ }(\omega^{ }_{q_j})}{q^2 \omega^{ }_{q_j}}
  - 
  \frac{2}{(4\pi)^2} \ln\biggl( \frac{\bmu^2}{m_j^2} \biggr)
 \la{Phi6_res} \\ 
 &  & \; + \, 
  \frac{ m_j^2 }{2} 
  \biggl[ 
  \int_q \frac{\nF^{ }(\omega^{ }_{q_j})}{ q^2 \omega^{3}_{q_j}}
  + 
  \frac{1}{T}
  \int_q \frac{\nF^{ }(\omega^{ }_{q_j})
  [ 1 - \nF^{ }(\omega^{ }_{q_j}) ] }{ q^2 \omega^{2}_{q_j}}
 \biggr]
 \biggr\}
 \;, \nonumber \hspace*{4mm} \\ 
 \Phi^\rmii{(7)}_{ } 
  &
    \hspace*{-0.4cm}
    \stackrel{m^{ }_h \!\ll m^{ }_t}{\approx}
    \hspace*{-0.4cm} 
  & 
 - \frac{ m_t^2 }{2} \biggl\{ 
  \int_p \frac{\nF^{ }(\omega^{ }_{p_t})}{ p^2 \omega^{3}_{p_t}}
  + 
  \frac{1}{T}
  \int_p \frac{\nF^{ }(\omega^{ }_{p_t})
  [ 1 - \nF^{ }(\omega^{ }_{p_t}) ] }{ p^2 \omega^{2}_{p_t}}
 \biggr\} 
 \la{Phi7_res} \\ 
 &  & \; \times \, 
 \biggl\{
  \frac{T^2}{12} 
  - \frac{3 m_t^2}{(4\pi)^2}
  \biggl[ 
      \ln\biggl(\frac{\bmu^2}{m_t^2}\biggr) 
    + \frac{7}{3} 
  \biggr]
 + 
 \int_q \frac{\nF^{ }(\omega^{ }_{q_t})}{\omega^{ }_{q_t}}
 \biggl( 1 + \frac{2 m_t^2}{q^2} \biggr)
 \biggr\}
 \;, \nonumber 
\ea
where $m^{ }_i$ refer to $\msbar$ masses evaluated at the 
renormalization scale $\bmu$.
In the limit $m^{ }_i \ll T$ the coefficients go over into 
\ba
 \Phi^\rmii{(2)}_{i}  & \stackrel{m^{ }_{i} \ll T}{\approx} &  
 \frac{T^2}{6} 
 \;, \la{Phi2_lim} \\ 
 \Phi^\rmii{(4)}_{i}  & \stackrel{m^{ }_{i} \ll T}{\approx} &  
 \frac{7T^2}{9(4\pi)^2} \biggl[ 
   \ln\biggl(\frac{\bmu e^{\gammaE}}{4 \pi T} \biggr)
 - \frac{8\ln 2}{7} + \frac{9}{14} \biggr]
 \;, \la{Phi4_lim} \\ 
 \Phi^\rmii{(5)}_{i}  & \stackrel{m^{ }_{i} \ll T}{\approx} &  
 - \frac{T^2}{(4\pi)^2} 
 \;, \la{Phi5_lim} \\ 
 \Phi^\rmii{(6)}_{j}  & \stackrel{m^{ }_{j} \ll T}{\approx} &  
 - \frac{4}{3(4\pi)^2}
 \biggl[ 
  \ln\biggl(\frac{\bmu e^{\gammaE}}{\pi T} \biggr) - \frac{1}{2}
 \biggr]
 \;, \la{Phi6_lim} \\ 
 \Phi^\rmii{(7)}_{ }  & \stackrel{m^{ }_{t,h} \ll T}{\approx} &  
 - \frac{T^2}{4(4\pi)^2} 
 \;, \la{Phi7_lim} 
\ea
reproducing in the first four cases the expressions
obtained in ref.~\cite{bn}. For $m^{ }_i \gg \pi T$, 
all terms containing $\nF^{ }$ are exponentially suppressed. 

\begin{figure}[t]

\hspace*{-0.1cm}
\centerline{%
 \epsfysize=7.6cm\epsfbox{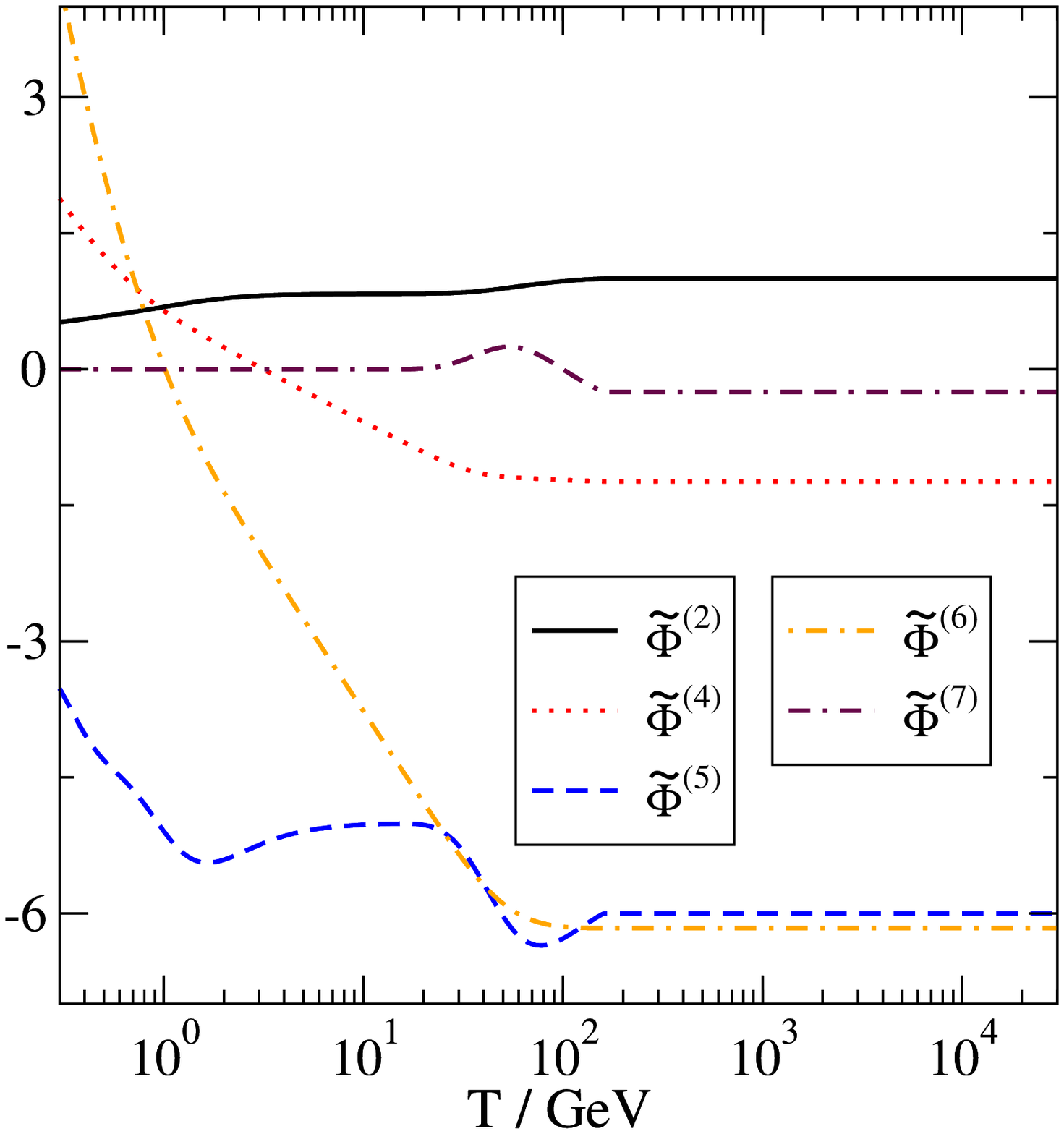}%
 \hspace*{7mm}\epsfysize=7.6cm\epsfbox{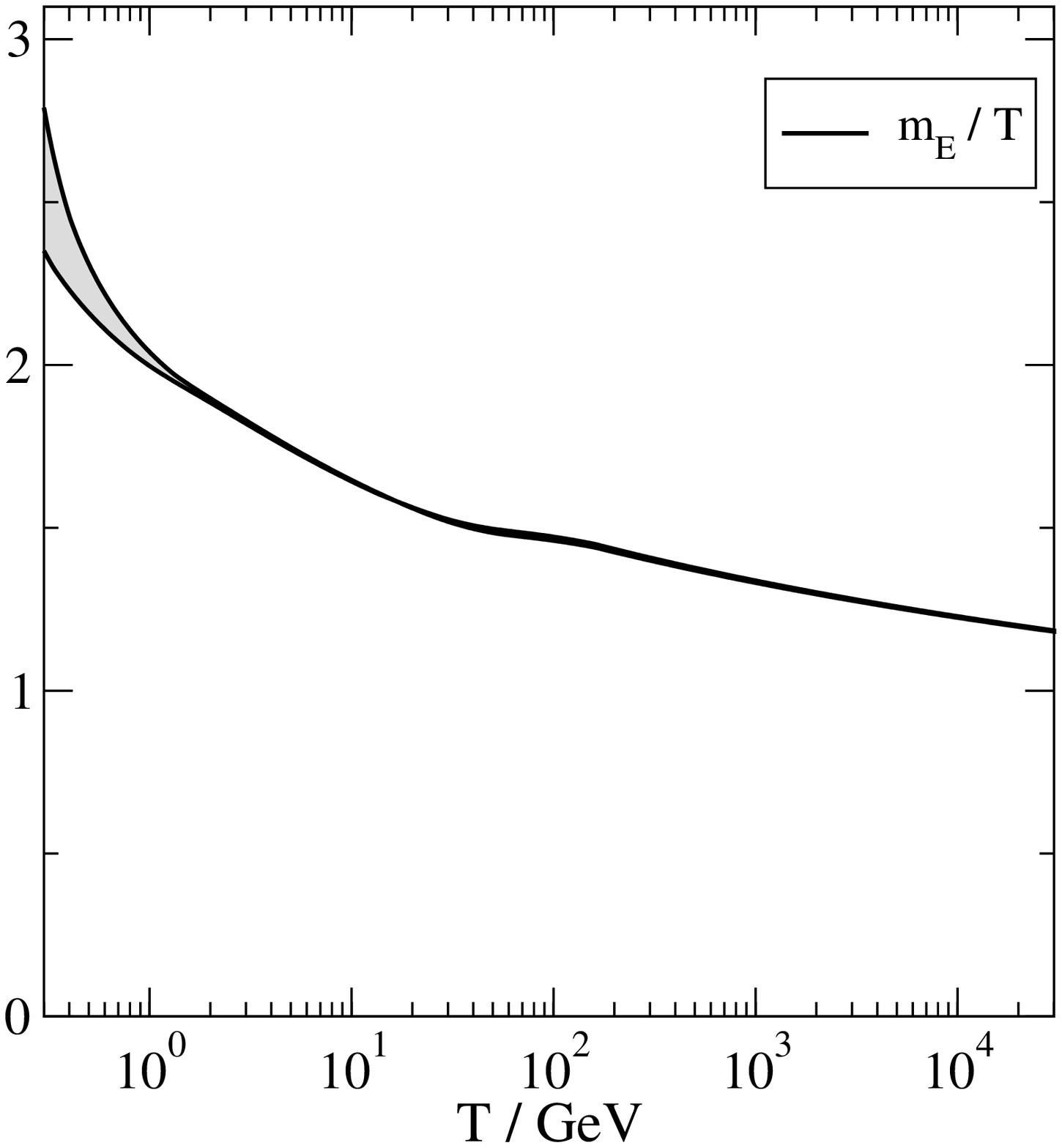}
}

\caption[a]{\small
 Left: the coefficients
 $
  \widetilde{\Phi}^{(2)}_{ } 
     \; \equiv \; 
  \sum_{i=1}^{\Nf^{ }} \Phi^{(2)}_i/T^2
 $, 
 $
  \widetilde{\Phi}^{(4,5)}_{ } 
     \; \equiv \; 
  \sum_{i=1}^{\Nf^{ }} (4\pi)^2 \Phi^{(4,5)}_i/T^2
 $, 
 $
  \widetilde{\Phi}^{(6)}_{ } 
     \; \equiv \; 
  \sum_{j=1}^{\Nf^{ }} (4\pi)^2 \Phi^{(6)}_{j}
 $ 
 and 
 $
  \widetilde{\Phi}^{(7)}_{ } 
     \; \equiv \; 
 (4\pi)^2 \Phi^{(7)}_{ }/T^2
 $ 
 that parametrize \eq\nr{Phi_def}, 
 evaluated with $\bmu = 2\pi T$. 
 Right: the QCD Debye mass as a function of the temperature.
 They grey band originates from varying the renormalization
 scale in the range $\bmu = (0.5 ... 2.0) \times 2\pi T$ and gives
 an indication of the magnitude of higher-order corrections.  
 The ``hard scale'' with which $\mE^{ }$ can be 
 compared is $\sim 2\pi T$. 
 The plateau-like feature
 centered around 
 $T \sim 70$~GeV originates 
 from crossing the top mass threshold. 
}

\la{fig:mmEoT2}
\end{figure}

For a numerical evaluation, 
we set $\alphas^{ }(\mZ^{ }) \simeq 0.118$~\cite{pdg}, 
and evolve $g^2(\bmu)$ in both directions with 5-loop 
running~\cite{alph1,alph2,alph3,alph4}, changing $\Nf^{ }$ when
a threshold is crossed at $\bmu \simeq m^{ }_i$, 
and including effects from the top Yukawa 
up to 3-loop order~\cite{max,mss}. Quark masses 
are likewise evolved at 5-loop level~\cite{mass1,mass2},
including effects from the top Yukawa as indicated below
\eq\nr{PiE1_3}.\footnote{%
  Running quark masses become ambiguous at $\bmu \gsim \mZ^{ }$,
  given that corrections to the Higgs vacuum expectation value from
  weak interactions are partly gauge dependent.
  That said, the SU$^{ }_\rmii{L}$(2)$\times$U$^{ }_\rmii{Y}$(1) 
  gauge effects are numerically very small compared
  with QCD corrections, 
  as already alluded to in footnote~\ref{yukawa},
  so we do not enter a more detailed discussion of this topic here. 
  } 
In order to account for temperature-dependent
tadpole corrections $\propto T^2$, 
the Higgs expectation value and $m^{ }_i$ are further scaled as 
$v^{ }_\T \simeq v^{ }_0 \, \re \sqrt{1 - (T/160~\mbox{GeV})^2} $
where $v^{ }_0 \simeq 246$~GeV and the crossover temperature
has been adopted from ref.~\cite{dono} rather
than from a perturbative computation.
As the top quark Yukawa coupling plays a minor role, we have
resorted to 2-loop running for $h_t^2$, with the initial
condition $h_t^2(\mZ^{ }) \simeq 0.95$ and running taking place
only for $\bmu > \mZ^{ }$.  The initial value of 
the running top mass
(before applying thermal rescaling) is estimated as 
$m_t(\mZ^{ }) \simeq h^{ }_t(\mZ^{ })v^{ }_0/\sqrt{2} \simeq 169.5$~GeV, 
whereas those of the quartic Higgs and electroweak couplings,
which affect the running
of $h_t^2$, are $\lambda (\mZ^{ }) \simeq 0.145$,
$g_1^2(\mZ^{ }) \simeq 0.128$,
$g_2^2(\mZ^{ }) \simeq 0.425$, respectively.
The renormalization scale is set to 
$\bmu = (0.5 ... 2.0) \times 2\pi T$, 
with the variation providing an error band. 
The results are plotted in \fig\ref{fig:mmEoT2}.

%
\section{Conclusions and outlook} 
\la{se:concl}

The goal of this technical contribution,  motivated by the
potential cosmological applications mentioned in \se\ref{se:intro}, 
has been to estimate
a QCD Debye mass, defined as a matching coefficient of the 
dimensionally reduced effective theory, at temperatures between
1~GeV and 10~TeV. For this purpose we have carried out a 2-loop computation, 
reducing the result to a small number of exponentially convergent
1- and 2-dimensional integrals, which are readily evaluated numerically. 

The most non-trivial parts of our work established
the IBP relations in \eqs\nr{red1}--\nr{red3} 
and resolved the
2-loop master sum-integral $Z^{ }_{112}$ in appendix~\ref{ss:z112}.
With these ingredients, we obtain
integral representations for the various functions 
parametrizing our result, cf.\ \eq\nr{Phi_def}, which are shown 
in \eqs\nr{Phi2_res}--\nr{Phi7_res} and evaluated
numerically in \fig\ref{fig:mmEoT2}(left). 
Putting everything together and inserting the values of 
running Standard Model couplings, we find that 
quark mass thresholds
are crossed smoothly enough not to be discernible by bare eye, apart from
that related to the top quark, cf.\ \fig\ref{fig:mmEoT2}(right). 

The steps of
the computation have been described on a detailed level, in order to
permit for the inclusion of further massive particles if present,
such as of scalar fields. 
Hopefully these results or techniques
can find use e.g.\ in dark matter computations involving strongly interacting 
co-annihilation partners, or in precision studies of the electroweak 
phase transition in extensions of the Standard Model. 

%
\section*{Acknowledgements}

This work was partly supported by the Swiss National Science Foundation
(SNF) under grant 200020B-188712
and by the Chilean FONDECYT under project 1191073.

%
\appendix
\renewcommand{\thesection}{Appendix~\Alph{section}}
\renewcommand{\thesubsection}{\Alph{section}.\arabic{subsection}}
\renewcommand{\theequation}{\Alph{section}.\arabic{equation}}

%
\section{Master sum-integrals} 

We list here the expressions for the master 
sum-integrals appearing in \eqs\nr{Phi1}--\nr{Phi7}. 

%
\subsection{1-loop structures} 

We start by reiterating the expressions for 
a number of 1-loop master sum-integrals, defined according
to \eq\nr{Z1l}. 
General techniques for evaluating massless sum-integrals 
were developed in refs.~\cite{az1,az2}. In the bosonic case, 
\be
 \bZ^{ }_{1;0} = \frac{T^2}{12}
 \, \biggl\{
   1 + 2\epsilon \, \biggl[
   \ln\biggl( \frac{\bmu e^{\gammaE}}{4\pi T} \biggr)
    + \ln(2\pi) - (\ln\zeta^{ }_2)'
  \biggr] 
  + \rmO(\epsilon^2)
 \biggr\}
 \;, 
\ee
where $\zeta^{ }_n = \zeta(n)$ is the Riemann zeta function,
$(\ln \zeta^{ }_n)' \equiv \zeta'(n)/\zeta(n)$,
and $\bmu^2 \equiv 4\pi \mu^2 e^{-\gammaE}$. 
In the literature a different form is often shown, 
obtained by employing the identity
$
 \ln(2\pi) - (\ln\zeta^{ }_2)' = 
  1 - \gammaE^{ } + (\ln\zeta^{ }_{-1})'
$.
A quadratic propagator similarly yields
\ba
 \bZ^{ }_{2;0} & = & 
 \frac{1}{(4\pi)^2} \biggl\{ \frac{1}{\epsilon}
 + 2 \ln \biggl( \frac{\bmu e^{\gammaE}}{4\pi T} \biggr)
 +   2\epsilon\, \biggl[ 
 \ln^2 \biggl( \frac{\bmu e^{\gammaE}} {4\pi T} \biggr)
 + \frac{\pi^2}{8} - \gammaE^2 - 2 \gamma^{ }_1
   \biggr]
 + \rmO(\epsilon^2)  
 \biggr\} 
 \;, \hspace*{5mm}
\ea
where $\gamma^{ }_1$ is a Stieltjes constant. More generally, 
$
 \bZ^{ }_{j;0} = 
 \frac{\bmu^{3-d} \exp[(3-d)\gammaE/2] \Gamma(j-{d}/{2})\zeta(2j-d)}
      {8 \pi^{{5}/{2}}(2\pi T)^{2j-1-d}\Gamma(j)}
$.

In the fermionic case, when the mass is non-zero, 
no analytic expressions are available. Even if convergent sum
representations in terms of modified Bessel functions can be found, 
in practice it is simpler to handle integral representations, 
such as 
\ba
 && \hspace*{ -1.0cm} Z^{ }_{1;i} \; = \; 
 - \frac{m_i^2}{(4\pi)^2\epsilon}
 - \frac{m_i^2}{(4\pi)^2} \biggl[ 
   \ln\biggl( \frac{\bmu^2}{m_i^2} \biggr) +1 \biggr]
 - \int_p \frac{\nF^{ }(\omega^{ }_{p_i})}{\omega^{ }_{p_i}}
 \\
 \!\! & - & \!\! 
 \epsilon 
 \, \biggl\{ 
  \frac{m_i^2}{(4\pi)^2}
  \biggl[
    \frac{1}{2} \ln^2 \biggl( \frac{\bmu^2}{m_i^2} \biggr)  
  + \ln \biggl( \frac{\bmu^2}{m_i^2} \biggr)
  + 1 + \frac{\pi^2}{12}
  \biggr]
  + 
  \int_p \frac{\nF^{ }(\omega^{ }_{p_i})}{\omega^{ }_{p_i}}
  \, \biggl[
       \ln \biggl( \frac{\bmu^2}{4p^2} \biggr) + 2  
  \biggr]\; 
 \biggr\} 
 + \rmO(\epsilon^2)
 \;, \nonumber  
\ea
where 
$
 \int_p \equiv \int \! \frac{{\rm d}^3\vec{p}}{(2\pi)^3}
$
and 
$
 \omega^{ }_{p_i} \equiv \sqrt{p^2 + m_i^2}
$.
Taking a mass derivative and carrying out a 
partial integration gives
\ba
 Z^{ }_{2;i} & = & \frac{1}{(4\pi)^2\epsilon}
 + \frac{1}{(4\pi)^2} 
   \ln\biggl( \frac{\bmu^2}{m_i^2} \biggr)
 - \int_p \frac{\nF^{ }(\omega^{ }_{p_i})}{2 p^2 \omega^{ }_{p_i}}
 \nn 
 & + & 
 \epsilon 
 \, \biggl\{ 
  \frac{1}{(4\pi)^2}
  \biggl[
    \frac{1}{2} \ln^2 \biggl( \frac{\bmu^2}{m_i^2} \biggr)  
  + \frac{\pi^2}{12}
  \biggr]
  - 
  \int_p \frac{\nF^{ }(\omega^{ }_{p_i})}{2 p^2 \omega^{ }_{p_i}}
       \ln \biggl( \frac{\bmu^2}{4p^2} \biggr)
 \biggr\} 
 + \rmO(\epsilon^2)
 \;. 
\ea
One more mass derivative yields (this time no partial
integration is possible; $\beta \equiv 1/T$) 
\ba
 Z^{ }_{3;i} & = & 
 \frac{1}{(4\pi)^2 2m_i^2} 
 - \int_p \frac{\nF^{ }(\omega^{ }_{p_i}) + 
   \beta \omega^{ }_{p_i} \nF^{ }(\omega^{ }_{p_i}) 
   [ 1 - \nF^{ }(\omega^{ }_{p_i}) ]}{8 p^2 \omega^{3}_{p_i}}
 \nn 
 & + & 
 \epsilon 
 \, \biggl\{ 
  \frac{
   \ln(\bmu^2/m_i^2) 
    }{(4\pi)^2 2 m_i^2}        
  - 
  \int_p \frac{\nF^{ }(\omega^{ }_{p_i}) + 
   \beta \omega^{ }_{p_i} \nF^{ }(\omega^{ }_{p_i}) 
   [ 1 - \nF^{ }(\omega^{ }_{p_i}) ] }{p^2 \omega^{3}_{p_i}}
       \ln \biggl( \frac{\bmu^2}{4p^2} \biggr)
 \biggr\} 
 + \rmO(\epsilon^2)
 \;. \hspace*{7mm}
\ea

%
\subsection{2-loop master $Z^{ }_{112}$} 
\la{ss:z112}

Even if in the massless limit IBP identities allow to reduce
$Z^{ }_{112}$ as
\be
 \lim_{m \to 0} Z^{ }_{112}
 = 
 \lim_{m^{ }_i\to 0}  
 \frac{
   Z^{ }_{2;i}(Z^{ }_{2;i} - 2 \bZ^{ }_{2;0})
 }
 {(d-2)(d-5)} 
 \;, \la{z112_massless} 
\ee
no such factorization has been found for $m\neq 0$. 
The result for a fully massive $Z^{ }_{111}$ is given 
in ref.~\cite{veff}, and one might think that $Z^{ }_{112}$
could be obtained as a mass derivative thereof, however this does not
work trivially as setting the third mass to zero after the derivative
leads to IR divergences (linear and logarithmic). 
A careful consideration is thus needed for working out the reduction of 
$Z^{ }_{112}$ into a convergent 2-dimensional integral representation.

As a first step, let us carry out the Matsubara sums. The quadratic
propagator carries a fictitious mass parameter, 
denoted by $M^2$, as an intermediate regulator. 
The sum-integral splits into a vacuum part,
one-cut parts, and two-cut parts, with ``cut'' meaning that
some line is put on-shell and weighted by a thermal distribution: 
\ba
 Z^{ }_{112} & = & 
 Z^\rmi{(vac)}_{112} + 
 Z^\rmii{(B)}_{112} + 
 Z^\rmii{(F)}_{112} + 
 Z^\rmii{(FB)}_{112} + 
 Z^\rmii{(FF)}_{112}
 \;, \la{z112_splitup} \\[2mm] 
 Z^\rmi{(vac)}_{112} & = & 
 \int_{P,Q} \frac{1}{(P^2+m^2)(Q^2+m^2)(P-Q)^4} 
 \;, \\[2mm] 
 Z^\rmii{(B)}_{112}
 & = & 
 - \lim_{M\to 0}
 \frac{{\rm d}}{{\rm d}M^2}
 \int_p \frac{\nB^{ }(\Omega^{ }_p)}{\Omega^{ }_p}
 \biggl[ 
  \int_Q \frac{1}{(Q^2+m^2)[(P-Q)^2+m^2]}
 \biggr]^{ }_{P^2 = - M^2}
 \;, \\[2mm] 
 Z^\rmii{(F)}_{112}
 & = & 
 - 2 \lim_{M\to 0}
 \int_p \frac{\nF^{ }(\omega^{ }_p)}{\omega^{ }_p}
 \biggl[ 
  \int_Q \frac{1}{(Q^2+M^2)^2[(P-Q)^2+m^2]}
 \biggr]^{ }_{P^2 = - m^2}
 \;, \\[2mm] 
 Z^\rmii{(FB)}_{112}
 & = & 
 2 \lim_{M\to 0}
 \frac{{\rm d}}{{\rm d}M^2}
 \int_{p,q} \frac{\nF^{ }(\omega^{ }_p)\nB^{ }(\Omega^{ }_q)}
 { \omega^{ }_p \Omega^{ }_q}
 \biggl[ 
  \frac{1}{(P-Q)^2+m^2}
 \biggr]^{ }_{P^2 = -m^2, Q^2 = -M^2}
 \;, \\[2mm] 
 Z^\rmii{(FF)}_{112}
 & = & 
 \lim_{M\to 0}
 \int_{p,q} \frac{\nF^{ }(\omega^{ }_p)\nF^{ }(\omega^{ }_q)}
 {\omega^{ }_p \omega^{ }_q}
 \biggl[ 
  \frac{1}{[(P-Q)^2+M^2]^2}
 \biggr]^{ }_{P^2 = - m^2, Q^2 = -m^2}
 \;. 
\ea
Here
$\Omega^{ }_p \equiv \sqrt{p^2 + M^2}$,
$\omega^{ }_p \equiv \sqrt{p^2 + m^2}$,
$\int_P \equiv \mu^{3-d}\int \! \frac{{\rm d}^{d+1} P}{(2\pi)^{d+1}}$,
$\int_p \equiv \mu^{3-d}\int \! \frac{{\rm d}^d \vec{p}}{(2\pi)^d}$.
The cuts are 
\be
 [...]^{ }_{P^2 = -m^2} \equiv \frac{1}{2} 
 \sum_{p^{ }_n = \pm i \omega^{ }_p} [...]
 \;, \quad 
 [...]^{ }_{P^2 = -m^2,Q^2=-M^2} \equiv \frac{1}{4} 
 \sum_{p^{ }_n = \pm i \omega^{ }_p} 
 \sum_{q^{ }_n = \pm i \Omega^{ }_q} 
 [...]
 \;. 
\ee
The parameter $M$ is set to zero for 
the final spatial integrals, which are treated with strict dimensional
regularization (this is necessary, given that the IBP identities used 
for reducing the result to this basis made use of the same recipe). 

Two of the structures in \eq\nr{z112_splitup} are simple to handle, 
namely the vacuum part and the one-cut part with a single Bose
distribution: 
\ba
 Z_{112}^\rmi{(vac)} & \stackrel{d=3-2\epsilon}{=} &
\frac{1}{(4\pi)^4}
 \biggl\{ 
   - \frac{1}{2\epsilon^2}
   + \frac{1}{\epsilon}\,
     \biggl[ \frac{1}{2} - 
            \ln\biggl( \frac{\bmu^2}{m^2} \biggr)\biggr]
   - \ln^2\biggl( \frac{\bmu^2}{m^2} \biggr)
   + \ln\biggl( \frac{\bmu^2}{m^2} \biggr) 
   - \frac{3}{2} - \frac{\pi^2}{12}
 \biggr\} 
 \;, \nn \la{z112_vac} \\
 Z_{112}^\rmii{(B)} & \stackrel{d=3-2\epsilon}{=} &
 - \frac{ \bZ^{ }_{1;0} }{6m^2 (4\pi)^2}
 + \frac{ \bZ^{ }_{2;0} }{(4\pi)^2}
 \biggl\{ 
  \frac{1}{\epsilon} 
 + 
  \ln\biggl( \frac{\bmu^2}{m^2} \biggr) 
 + 
  \epsilon \, 
  \biggl[
   \frac{1}{2} 
   \ln^2\biggl( \frac{\bmu^2}{m^2} \biggr) 
   + \frac{\pi^2}{12}
  \biggr] 
 \biggr\} 
 \;. \la{z112_B} 
\ea 
The remaining parts are more subtle, as they are IR divergent, in 
a way which is not trivially handled by dimensional regularization. 

Let us start by considering $ Z_{112}^\rmii{(FF)} $, which
contains a linear IR divergence but no logarithmic one. 
As this integral is UV finite, we may set $d=3$ and 
carry out the angular integral, which yields
\be
 Z^\rmii{(FF)}_{112} \stackrel{d=3}{\simeq} 
 \frac{1}{4m^2}
 \int_{p,q} 
 \frac{\nF^{ }(\omega^{ }_p)\nF^{ }(\omega^{ }_q)}
 {\omega^{ }_p \omega^{ }_q} 
 \frac{\omega^{2}_p + \omega^{2}_q}
 {(\omega^{2}_p - \omega^{2}_q)^2}
 \;. 
\ee
Clearly this is ill-defined around $p = q$. 
In order to find a useful representation, we make use of 
symmetries of the integrand, re-organizing the Fermi 
distributions as 
\ba
  Z^\rmii{(FF)}_{112} & \stackrel{d=3}{=} & 
 \frac{1}{16m^2}
 \int_{p,q} 
 \frac{1}
 {\omega^{ }_p \omega^{ }_q} 
 \biggl\{ 
   \frac{
    [ \nF^{ }(\omega^{ }_p) + \nF^{ }(\omega^{ }_q) ]^2
        }{
    ( \omega^{ }_p + \omega^{ }_q )^2
        }
   - 
   \frac{
    [ \nF^{ }(\omega^{ }_p) - \nF^{ }(\omega^{ }_q) ]^2
        }{
    ( \omega^{ }_p - \omega^{ }_q )^2
        }
 \biggr\} 
 + \delta Z^\rmii{(FF)}_{112}
 \;, \nn \la{z112_FF} \\  
 \delta Z^\rmii{(FF)}_{112}
 & \stackrel{d=3}{=} & 
 \int_p \frac{\nF^2(\omega^{ }_p)}{2m^2}
 \int_q \frac{1}{(q^2 - p^2)^2}
 \;.  \la{delta_z112_FF}
\ea
Considering  
the vacuum-like integral in \eq\nr{delta_z112_FF}
as an analytic function of $-p^2$
and taking the real part, yields
$
 \re \int_q \frac{1}{(q^2 - p^2)^2} = 
 \re \frac{1}{8\pi (-p^2)^{1/2}}
  = 0
$.
Alternatively, if we keep the regulator $M$ finite, 
$ \delta Z^\rmii{(FF)}_{112} $ contains a linear
divergence\footnote{%
  The sum of all $1/M$-divergences in $Z^{ }_{112}$
  equals the Matsubara zero-mode contribution
  $
   T\int_p \frac{1}{(p^2 + M^2)^2} \, Z^{ }_{2;i}
  $.
}
$\propto 1/M$ but no logarithmic 
or finite part of $\rmO(M^0)$. To summarize, 
in strict dimensional regularization we can set 
$ \delta Z^\rmii{(FF)}_{112} \to 0$.

The remaining parts, 
$  Z^\rmii{(F)}_{112} $  
and 
$ Z^\rmii{(FB)}_{112} $, 
contain both linear and logarithmic divergences. 
The logarithmic divergences cancel in the sum. 
We find it practical to determine the sum by keeping
$M$ finite and taking $M\to 0$ at the end, 
omitting again linear divergences $\propto 1/M$, 
which are absent in strict dimensional regularization.
A rather tedious analysis then yields 
\be
 Z^\rmii{(F)}_{112} 
 + 
 Z^\rmii{(FB)}_{112} 
 \; \stackrel{d=3}{=} \; 
 \frac{2}{m^2(4\pi)^2}
 \int_p \! \frac{\nF^{ }(\omega^{ }_p)}{\omega^{ }_p}
 \biggl[
  \ln\biggl( \frac{m e^{\gammaE}}{4\pi T} \biggr) 
 + 1 
 + \frac{\omega^{ }_p}{2p}
   \ln\biggl(\frac{ \omega^{ }_p + p }{\omega^{ }_p - p}\biggr)
 \biggr]
 \;. \la{z112_F_BF}
\ee

Given the non-triviality of the steps, it is
good to check that \eq\nr{z112_massless} is correctly reproduced
for $m/T\to 0$. 
The individual parts contain coefficients $\propto 1/m^2$, 
so we need to expand to $\rmO(m^2)$. 
The integral appearing in \eq\nr{z112_F_BF} can be expanded as
\ba
 && \hspace*{-1.5cm}
 \int_p \! \frac{\nF^{ }(\omega^{ }_p)}{\omega^{ }_p}
 \biggl[
  \ln\biggl( \frac{me^{\gammaE}}{4\pi T} \biggr) 
 + 1 
 + \frac{\omega^{ }_p}{2p}
   \ln\biggl(\frac{ \omega^{ }_p + p }{\omega^{ }_p - p}\biggr)
 \biggr]
 \; = \; 
 \frac{T^2}{24} \, 
 \bigl[
   2 + (\ln \zeta^{ }_2)' - \ln\pi  
 \bigr]
 \nn 
 & + & 
 \frac{2m^2}{(4\pi)^2}
 \biggl[
   \ln^2\biggl( \frac{m e^{\gammaE}}{4\pi T} \biggr) 
  + (1 + 2 \ln 2)
   \ln\biggl( \frac{m e^{\gammaE}}{4\pi T} \biggr) 
  + 3\ln2 - \frac{1}{2}  
 \biggr]
 + \rmO(m^4)
 \;, 
\ea
whereas the contribution from \eq\nr{z112_FF} 
can be numerically verified to behave as 
\ba
 & & \hspace*{-1.5cm} 
 \int_{p,q} 
 \frac{1}
 {\omega^{ }_p \omega^{ }_q} 
 \biggl\{ 
   \frac{
    [ \nF^{ }(\omega^{ }_p) + \nF^{ }(\omega^{ }_q) ]^2
        }{
    ( \omega^{ }_p + \omega^{ }_q )^2
        }
   - 
   \frac{
    [ \nF^{ }(\omega^{ }_p) - \nF^{ }(\omega^{ }_q) ]^2
        }{
    ( \omega^{ }_p - \omega^{ }_q )^2
        }
 \biggr\} 
 = 
 - \frac{4T^2}{3(4\pi)^2} \, 
 \biggl[
   \frac{11}{6} + (\ln \zeta^{ }_2)' - \ln\pi  
 \biggr]
 \nn  
 & - & 
 \frac{32m^2}{(4\pi)^4}
 \biggl[
   \ln^2\biggl( \frac{m e^{\gammaE}}{\pi T} \biggr) 
  + \ln\biggl( \frac{m e^{\gammaE}}{\pi T} \biggr) 
  + 4\ln2 - \frac{5}{2}  
 \biggr]
 + \rmO(m^4)
 \;. 
\ea
Summing together and 
adding the other parts, 
we reproduce the result from 
\eq\nr{z112_massless},
\ba 
 \lim_{m^{ } \to 0} Z^{ }_{112}
 & \stackrel{d=3-2\epsilon}{=} & 
 \frac{1}{(4\pi)^4}
 \biggl\{ 
   \frac{1}{2\epsilon^2}  
  + \frac{1}{\epsilon}\,
    \biggl[ 
     \frac{1}{2} + 
  2 \ln\biggl( \frac{\bmu e^{\gammaE}}{4\pi T} \biggr)
    \biggr]
 + 
 4 \ln^2\biggl( \frac{\bmu e^{\gammaE}}{4\pi T} \biggr)
 + 
 2 \ln\biggl( \frac{\bmu e^{\gammaE}}{4\pi T} \biggr)
 \nn 
 &  & \quad 
 - \,
 8 \ln^2 2 
 + \frac{\pi^2}{4}
 + \frac{3}{2} 
 - 2 \gammaE^2
 - 4 \gamma^{ }_1
 \biggr\} 
 \;.
\ea

%
\subsection{2-loop master $Z^{ }_{212}$} 

In the massless limit IBP identities allow to reduce $Z^{ }_{212}$ as
\be
 \lim_{m^{ } \to 0} Z^{ }_{212}
 = 
 \lim_{m^{ }_i \to 0}
 \frac{
  2 Z^{ }_{3;i}  
   (Z^{ }_{2;i} - \bZ^{ }_{2;0})
 }
 {(d-2)(d-7)} 
 \;. \la{z212_massless} 
\ee
For a finite mass, we can instead write
\be
 Z^{ }_{212} = - \frac{1}{2} \frac{{\rm d} Z^{ }_{112} }{{\rm d}m^2}
 \;. 
\ee
Converting a number of mass derivatives into 
derivatives with respect to momentum, 
and carrying out partial integrations, 
\eqs\nr{z112_vac}, 
\nr{z112_B}, \nr{z112_FF} and \nr{z112_F_BF} then imply that
\ba
 Z^{ }_{212} & \stackrel{d=3-2\epsilon}{=} & 
 - \frac{1}{2m^2(4\pi)^4}
 \biggl[
  \frac{1}{\epsilon} 
  + 2 \ln\biggl( \frac{\bmu^2}{m^2} \biggr)
  -1 
 \biggr]
 \nn 
 & - &  
 \frac{ \bZ^{ }_{1;0} }{12 m^4 (4\pi)^2}
 + 
 \frac{ \bZ^{ }_{2;0} }{2 m^2 (4\pi)^2}
 \biggl[ 1 + \epsilon 
      \ln\biggl( \frac{\bmu^2}{m^2} \biggr)
 \biggr]
 \nn 
 & + & 
 \frac{1}{m^4 (4\pi)^2}
 \int_p \! \frac{\nF^{ }(\omega^{ }_p)}{\omega^{ }_p}
 \biggl\{
  \biggl( 1 + \frac{m^2}{2p^2} \biggr)
  \ln\biggl( \frac{m e^{\gammaE}}{4\pi T} \biggr) 
 + \frac{\omega_p^2}{p^2}
 + \frac{\omega^{ }_p}{2p}
   \ln\biggl(\frac{ \omega^{ }_p + p }{\omega^{ }_p - p}\biggr)
 \biggr\}
 \nn 
 & + &
 \frac{1}{64m^4}
 \int_{p,q} 
 \frac{1}
 {\omega^{ }_p \omega^{ }_q} 
 \biggl( 
  \frac{\omega_p^2}{p^2}
 + 
  \frac{\omega_q^2}{q^2}
 \biggr)
 \biggl\{ 
   \frac{
    [ \nF^{ }(\omega^{ }_p) + \nF^{ }(\omega^{ }_q) ]^2
        }{
    ( \omega^{ }_p + \omega^{ }_q )^2
        }
   - 
   \frac{
    [ \nF^{ }(\omega^{ }_p) - \nF^{ }(\omega^{ }_q) ]^2
        }{
    ( \omega^{ }_p - \omega^{ }_q )^2
        }
 \biggr\}
 \;. \nn \la{z212_res}
\ea

An interesting crosscheck of \eq\nr{z212_res} can be 
obtained by considering the massless limit. As there are
coefficients $\propto 1/m^4$, we need to expand the 
integrals up to $\rmO(m^4)$,
\ba
 && \hspace*{-1.5cm}
 \int_p \! \frac{\nF^{ }(\omega^{ }_p)}{\omega^{ }_p}
 \biggl\{
  \biggl( 1 + \frac{m^2}{2p^2} \biggr)
  \ln\biggl( \frac{m e^{\gammaE}}{4\pi T} \biggr) 
 + \frac{\omega_p^2}{p^2}
 + \frac{\omega^{ }_p}{2p}
   \ln\biggl(\frac{ \omega^{ }_p + p }{\omega^{ }_p - p}\biggr)
 \biggr\}
 \; = \; 
 \frac{T^2}{24} \, 
 \bigl[
   2 + (\ln \zeta^{ }_2)' - \ln\pi  
 \bigr]
 \nn 
 & - & 
 \frac{2m^2}{(4\pi)^2}
 \biggl[
   \ln\biggl( \frac{m e^{\gammaE}}{4\pi T} \biggr) 
  + \ln2 + \frac{1}{2}  
 \biggr]
 + \frac{14 \zeta^{ }_3 m^4}{(4\pi)^4 T^2}
 \biggl[
   \ln\biggl( \frac{m e^{\gammaE}}{4\pi T} \biggr) + \frac{9}{4} 
 \biggr]
 + \rmO(m^6)
 \;, \\[2mm]
 && \hspace*{-1.5cm}
 \frac{1}{64}
 \int_{p,q} 
 \frac{1}
 {\omega^{ }_p \omega^{ }_q} 
 \biggl( 
  \frac{\omega_p^2}{p^2}
 + 
  \frac{\omega_q^2}{q^2}
 \biggr)
 \biggl\{ 
   \frac{
    [ \nF^{ }(\omega^{ }_p) + \nF^{ }(\omega^{ }_q) ]^2
        }{
    ( \omega^{ }_p + \omega^{ }_q )^2
        }
   - 
   \frac{
    [ \nF^{ }(\omega^{ }_p) - \nF^{ }(\omega^{ }_q) ]^2
        }{
    ( \omega^{ }_p - \omega^{ }_q )^2
        }
 \biggr\}
 \nn 
 & = & 
 - \frac{T^2}{24 (4\pi)^2} \, 
 \biggl[
   \frac{11}{6} + (\ln \zeta^{ }_2)' - \ln\pi  
 \biggr]
 \nn  
 & + & 
 \frac{m^2}{(4\pi)^4}
 \biggl[
   \ln\biggl( \frac{m e^{\gammaE}}{\pi T} \biggr) 
  + \frac{1}{2}  
 \biggr]
 - \frac{14 \zeta^{ }_3 m^4}{(4\pi)^6 T^2}
 \biggl[ 
  \ln\biggl( \frac{m e^{\gammaE}}{\pi T} \biggr) 
  + \frac{9}{4}
 \biggr]
 + \rmO(m^6)
 \;, 
\ea
the last of which was verified numerically. 
Summing together and 
adding the other terms, we recover 
the result from \eq\nr{z212_massless}, 
\be
 \lim_{m^{ }\to 0} Z^{ }_{212}
 \stackrel{d=3-2\epsilon}{=}
 -\frac{28 \zeta^{ }_3 \ln 2}{(4\pi)^6 T^2}
 \;. 
\ee

\small{
%

}

\end{document}